\documentclass
[
	aip,
	jcp,
	numerical,
	amsmath,
	amssymb,
reprint
]
{revtex4-1}

\usepackage[morefloats=20]{morefloats}

\usepackage{ifluatex}
\ifluatex

\else
\usepackage[T1]{fontenc}
\usepackage[utf8]{inputenc}
\DeclareUnicodeCharacter{2010}{-}
\fi

\usepackage{import}
\usepackage{xparse}

\usepackage{enumitem}

\usepackage[normalem]{ulem}

\usepackage{comment}

\usepackage{currfile}
\usepackage{float}
\usepackage{xstring}

\usepackage{pgfplots}
\usepackage{tikz}

\pgfplotsset{compat=1.3}

\usetikzlibrary{external}
\tikzset{external/system call={lualatex \tikzexternalcheckshellescape -shell-escape -halt-on-error -interaction=nonstopmode -jobname "\image" "\texsource"}}
\tikzset{external/up to date check=simple}
\tikzset{external/mode=list and make}

\NewDocumentCommand \tikzFigure { m o m o }%
{%
	\begin{figure}[H]\centering
		\IfNoValueF{#4}{\tikzsetnextfilename{#4}}
		\input{#1}
		\caption{%
			\IfNoValueF{#2}{\label{#2}}%
			(color online)
			#3%
			}
	\end{figure}
}

\usepackage{environ}
\usepackage{mathtools}
\usepackage{nicefrac}

\RenewDocumentCommand \cos { g }%
{%
	\IfNoValueTF{#1}
		{\ensuremath{\operatorname{cos}}}
		{\ensuremath{\operatorname{cos} \left(#1\right)}}%
}
\RenewDocumentCommand \cosh { g }%
{%
	\IfNoValueTF{#1}
		{\ensuremath{\operatorname{cosh}}}
		{\ensuremath{\operatorname{cosh} \left(#1\right)}}%
}

\newcommand{\differential}[0]{\ensuremath{\mathrm{d}}}
\NewDocumentCommand \diracDelta { g }%
{%
	\IfNoValueTF{#1}
		{\ensuremath{\delta}}
		{\ensuremath{\delta \left(#1\right)}}%
}

\NewEnviron{eq}{\begin{align} \BODY \end{align}}
\NewDocumentEnvironment{eqs}{ o }
{%
	\subequations
	\IfNoValueF{#1}{ \label{#1} }
	\align%
}
{%
	\endalign
	\endsubequations
	\ignorespacesafterend
}
\RenewDocumentCommand \exp { g }%
{%
	\IfNoValueTF{#1}
		{\ensuremath{\operatorname{exp}}}
		{\ensuremath{\operatorname{exp} \left(#1\right)}}%
}
\NewDocumentCommand \factorial { g }%
{%
	\IfNoValueTF{#1}
		{\ensuremath{!}}
		{\ensuremath{\left(#1\right)!}}%
}
\newcommand{\fourierDecoration}[1]{\ensuremath{\tilde{#1}}}
\NewDocumentCommand \GammaFunction { g }%
{%
	\IfNoValueTF{#1}
		{\ensuremath{\Gamma}}
		{\ensuremath{\Gamma\left(#1\right)}}%
}
\NewDocumentCommand \heavisideTheta { g }%
{%
	\IfNoValueTF{#1}
		{\ensuremath{\theta}}
		{\ensuremath{\theta \left(#1\right)}}%
}
\NewDocumentCommand \imaginaryPart { g }%
{%
	\IfNoValueTF{#1}
		{\ensuremath{\mathfrak{Im}}}
		{\ensuremath{\mathfrak{Im} \left(#1\right)}}%
}

\newcommand{\laplaceDecoration}[1]{\ensuremath{\hat{#1}}}

\NewDocumentCommand \postfixDifferential { m g }%
{%
	\IfNoValueTF{#2}
		{\ensuremath{\, \differential #1}}
		{\ensuremath{\, \differential^{#1} #2}}%
}

\NewDocumentCommand \realPart { g }%
{%
	\IfNoValueTF{#1}
		{\ensuremath{\mathfrak{Re}}}
		{\ensuremath{\mathfrak{Re} \left(#1\right)}}%
}
\RenewDocumentCommand \sin { o g }%
{%
	\IfNoValueTF{#2}
		{\ensuremath{\operatorname{sin}}}
		{\ensuremath{\operatorname{sin} \IfNoValueF{#1}{^{#1}} \left(#2\right)}}%
}
\RenewDocumentCommand \sinh { g }%
{%
	\IfNoValueTF{#1}
		{\ensuremath{\operatorname{sinh}}}
		{\ensuremath{\operatorname{sinh} \left(#1\right)}}%
}

\NewDocumentCommand \textfrac { o m o m }%
{%
	\IfNoValueTF{#1}
	{
		#2
	}
	{
		\left( #2 \right)
	}%
	/%
	\IfNoValueTF{#3}
	{
		#4
	}
	{
		\left( #4 \right)
	}%
}
\newcommand{\vect}[1]{\ensuremath{\boldsymbol{\mathbf{#1}}}}

\NewDocumentCommand \derivative { o m g }%
{%
	\IfNoValueTF{#1}
	{
		\frac{ \differential }{ \differential #2 }
	}
	{
		\frac{ \differential^{#1} }{ \differential {#2}^{#1} }	
	}%
	\IfNoValueF{#3}
	{
		\left( #3 \right)
	}%
}

\usepackage{csquotes}

\NewDocumentCommand \complexModulus { g }%
{%
	\IfNoValueTF{#1}
		{\ensuremath{G^*}}
		{\ensuremath{G^* \left(#1\right)}}%
}

\newcommand{\fluidDensity}[0]{\ensuremath{\varrho}}
\NewDocumentCommand \fluidVelocity { g g }%
{%
	\IfNoValueTF{#1}
		{\ensuremath{\vect{v}}}
		{\ensuremath{v_{#1}}}%
	\IfNoValueF{#2}{\left( #2 \right)}%
}
\NewDocumentCommand \fluidVelocityFT { g g }%
{%
	\IfNoValueTF{#1}
		{\ensuremath{\vect{\fourierDecoration{v}}}}
		{\ensuremath{\fourierDecoration{v}_{#1}}}%
	\IfNoValueF{#2}{\left( #2 \right)}%
}
\NewDocumentCommand \fluidVelocityFTT { g g }%
{%
	\IfNoValueTF{#1}
		{\ensuremath{\vect{\fourierDecoration{v}^{\textrm{T}}}}}
		{\ensuremath{\fourierDecoration{v}_{#1}^{\textrm{T}}}}%
	\IfNoValueF{#2}{\left( #2 \right)}%
}
\NewDocumentCommand \fourierVector { g g }%
{%
	\IfNoValueTF{#1}
		{\ensuremath{\vect{k}}}
		{\ensuremath{k_{#1}}}%
	\IfNoValueF{#2}{\left( #2 \right)}%
}
\NewDocumentCommand \fourierVectorIntegerComponent { g g }%
{%
	\IfNoValueTF{#1}
		{\ensuremath{\vect{k}_n}}
		{\ensuremath{k_{n, #1}}}%
	\IfNoValueF{#2}{\left( #2 \right)}%
}

\NewDocumentCommand \lossModulus { g }%
{%
	\IfNoValueTF{#1}
		{\ensuremath{G^{\prime\prime}}}
		{\ensuremath{G^{\prime\prime} \left(#1\right)}}%
}

\newcommand{\MPCParticleCenterOfMassVelocity}[1]{\ensuremath{\vect{v}_{\textrm{cm},i}}}

\NewDocumentCommand \position { g g }%
{%
	\IfNoValueTF{#1}
		{\ensuremath{\vect{r}}}
		{\ensuremath{r_{#1}}}%
	\IfNoValueF{#2}{\left( #2 \right)}%
}
\NewDocumentCommand \pressure { g }%
{%
	\IfNoValueTF{#1}
		{\ensuremath{p}}
		{\ensuremath{p \left( #1 \right)}}%
}
\NewDocumentCommand \pressureFT { g }%
{%
	\IfNoValueTF{#1}
		{\ensuremath{\fourierDecoration{p}}}
		{\ensuremath{\fourierDecoration{p} \left( #1 \right)}}%
}
\NewDocumentCommand \relaxationModulus { g }%
{%
	\IfNoValueTF{#1}
		{\ensuremath{G}}
		{\ensuremath{G \left(#1\right)}}%
}
\NewDocumentCommand \relaxationModulusLT { g }%
{%
	\IfNoValueTF{#1}
		{\ensuremath{\laplaceDecoration{G}}}
		{\ensuremath{\laplaceDecoration{G} \left(#1\right)}}%
}

\NewDocumentCommand \storageModulus { g }%
{%
	\IfNoValueTF{#1}
		{\ensuremath{G^\prime}}
		{\ensuremath{G^\prime \left(#1\right)}}%
}
\NewDocumentCommand \VACFFTT { g }%
{%
	\IfNoValueTF{#1}
		{\ensuremath{\fourierDecoration{C}_{\mathrm{v}}^{\mathrm{T}}}}
		{\ensuremath{\fourierDecoration{C}_{\mathrm{v}}^{\mathrm{T}} \left( #1 \right)}}%
}
\NewDocumentCommand \VACFFTTLT { g }%
{%
	\IfNoValueTF{#1}
		{\ensuremath{\laplaceDecoration{\fourierDecoration{C}}_{\mathrm{v}}^{\mathrm{T}}}}
		{\ensuremath{\laplaceDecoration{\fourierDecoration{C}}_{\mathrm{v}}^{\mathrm{T}} \left( #1 \right)}}%
}

\usepackage{catchfile}
\usepackage[exponent-product = \cdot]{siunitx}
\usepackage{trimspaces}

\usepackage{pdftexcmds}
\usepackage{xparse}

\makeatletter
\NewDocumentCommand{\assertEqual}{ m m  }
{
	\ifnum\pdf@strcmp{#1}{#2}=0
	\else
		\errmessage{
			Assertion failed in \assertEqual :
			"#1"
			does not equal the expected
			"#2"}
	\fi
}
\makeatother

\newcommand{\bulkThermostatTargetkTValue}{1.0}
\newcommand{\MPCCellLengthValue}{1}
\newcommand{\MPCParticleMassValue}{1}
\makeatletter
\trim@spaces@in{\bulkThermostatTargetkTValue}
\trim@spaces@in{\MPCCellLengthValue}
\trim@spaces@in{\MPCParticleMassValue}
\makeatother
\assertEqual{\bulkThermostatTargetkTValue}{1.0}
\assertEqual{\MPCCellLengthValue}{1}
\assertEqual{\MPCParticleMassValue}{1}

\newcommand{\defaultSimulationBoxSizeValue}{30}

\newcommand{\MPCAverageParticleCountPerCellValue}{10}
\newcommand{\MPCCollisionAngleValue}{2.27}
\newcommand{\MPCStreamingTimeStepValue}{0.1}

\makeatletter
\trim@spaces@in{\defaultSimulationBoxSizeValue}
\trim@spaces@in{\MPCAverageParticleCountPerCellValue}
\trim@spaces@in{\MPCCollisionAngleValue}
\trim@spaces@in{\MPCStreamingTimeStepValue}
\makeatother

\usepackage[capitalise]{cleveref}

\Crefname{equation}{Eq.}{Eqs.}
\Crefname{figure}{Fig.}{Figs.}
\Crefname{tabular}{Tab.}{Tabs.}

\providecommand{\onlinecite}{\cite}

\newcommand{\eqnref}[1]{\Cref{#1}}

\usepackage{pgfplots}
\usepackage{pgfplotstable}

\usepackage{etoolbox}
\usepackage{pdftexcmds}
\usepackage{xparse}
\usepackage{bm}

\newcommand{\lla}{\left\langle}
\newcommand{\rra}{\right\rangle}

\newcommand{\vGamma}{\varGamma}

\usepackage{grffile}
\graphicspath{{figures/}{tikzexternalize-cache/}}

\begin{document}

\title{Hydrodynamic correlations of viscoelastic fluids by multiparticle collision dynamics simulations}

\author{David \surname{Toneian}}
\email{david@toneian.com}

\affiliation{Institute for Theoretical Physics, TU Wien, Wiedner Hauptstraße 8-10, 1040 Wien, Austria}

\author{Gerhard \surname{Kahl}}
\email{gerhard.kahl@tuwien.ac.at}

\affiliation{Institute for Theoretical Physics, TU Wien, Wiedner Hauptstraße 8-10, 1040 Wien, Austria}
\affiliation{Center for Computational Materials Science (CMS), TU Wien, Wiedner Hauptstraße 8-10, 1040 Wien, Austria}

\author{Gerhard \surname{Gompper}}
\email{g.gompper@fz-juelich.de}

\author{Roland G. \surname{Winkler}}
\email{r.winkler@fz-juelich.de}

\affiliation{Theoretical Soft Matter and Biophysics, Institute of Complex Systems and Institute for Advanced Simulation, Forschungszentrum Jülich, 52425 Jülich, Germany}

\date{\today}

\begin{abstract}
The emergent fluctuating hydrodynamics of a viscoelastic fluid modeled by the multiparticle collision dynamics (MPC) approach is studied. The fluid is composed of flexible, Gaussian phantom polymers, which interact by local momentum-conserving stochastic MPC collisions. For comparison, the analytical solution of the linearized Navier-Stokes equation is calculated, where viscoelasticity is taken into account by a time-dependent shear relaxation modulus.  The fluid properties are  characterized by the transverse velocity autocorrelation function in Fourier space as well as in real space. Various polymer lengths are considered---from dumbbells to (near-)continuous polymers. Viscoelasticity affects the fluid properties and leads to strong correlations, which overall decay exponentially in Fourier space. In real space, the center-of-mass velocity autocorrelation function of individual polymers exhibits a long-time tail independent of polymer length, which decays as $t^{-3/2}$, similar to a Newtonian fluid, in the asymptotic limit $t \to \infty$. Moreover, for long polymers an additional power-law decay appears at time scales shorter than the longest polymer relaxation time with the same time dependence, but negative correlations, and the polymer length dependence $L^{-1/2}$. Good agreement is found between the analytical  and simulation results.
\end{abstract}

\maketitle

\section{Introduction} \label{sec:introduction}

Soft matter and complex fluids are composed of a broad range of nano- to microscale objects. Such systems are typically easily deformable, with characteristic energies on the order of the thermal energy and correspondingly long relaxation times,  and entropic degrees of freedom play an important role. \cite{dhon:08,menz:15,nage:17}  Paradigmatic examples of soft matter are biological cells---containing a wide range of polymeric and colloidal ingredients \cite{elli:01,bucc:16}---blood, solutions of polymers, emulsions, and suspensions of colloidal particles. \cite{bird:87.1,dhon:96} The majority of these suspensions are viscoelastic rather than Newtonian, combining the viscous properties of fluids with elastic characteristics of solids. \cite{ferr:80,bird:87,bird:87.1,doi:86,lars:99}

Computer simulations are a valuable tool for gaining insight into the viscoelastic properties of complex fluids. \cite{wink:14.1} Of particular interest are mesoscale simulation techniques, which account for hydrodynamic interactions and are able to bridge the length- and time-scale gap between fluid degrees of freedom and those of the embedded (polymeric) particles. \cite{tao:08,kowa:13} Established mesoscale techniques are the lattice Boltzmann method (LB), \cite{mcna:88,shan:93,succ:01,duen:09} dissipative particle dynamics (DPD), \cite{hoog:92,espa:95} and the multiparticle collision dynamics approach (MPC). \cite{male:99,kapr:08,gomp:09} Viscoelasticity is incorporated in different ways in the various simulation approaches. LB describes a fluid in terms of a spatially discretized probability density, whose dynamics progresses via the Boltzmann equation. \cite{succ:01,duen:09}  Viscoelasticity is incorporated by extending the stress tensor by a viscous-stress contribution, e.g., the Maxwell model, \cite{bird:87.1,ispo:02,dell:14} and taking this stress into account as a body force in the discretized propagation equation. \cite{ispo:02,dell:14} In contrast, DPD and MPC are particle-based simulation approaches, where the bare fluid is represented by point particles, and a complex fluid by additional  suspended objects such as colloids, polymers, membranes, or cells. In the latter approaches, viscoelasticity emerges as a consequence of the interactions between the embedded objects. Examples for viscoelastic  DPD  simulations are studies of blood cells \cite{fedo:11} and star polymers \cite{fedo:12} in flow. For MPC, the rheological properties of linear, branched, and star polymers \cite{huan:10,niko:10.1,sing:13,wink:13,fedo:12,tone:19,tone:19b} have been investigated, as well as that of cells and vesicles. \cite{nogu:04}

Alternatively,  viscoelastic fluids can be modeled by an ensemble of more complex entities, directly representing a viscoelastic fluid rather than a viscoelastic suspension. DPD and MPC viscoelastic fluids can be modeled by linearly connected DPD or MPC particles, respectively. The simplest viscoelastic unit is a dumbbell. The extension of the original DPD approach to a dumbbell fluid is presented in Ref.~\onlinecite{somf:06} and to even longer polymers in Ref.~\onlinecite{fedo:10}. Similarly, the properties of MPC dumbbell fluids of different complexity are studied  in Refs.~\onlinecite{tao:08,ji:11,kowa:13,tone:15}.

This representation of a viscoelastic fluid via an ensemble of linear elastic polymers raises a number of fundamental questions on hydrodynamic interactions in such a solution. Traditionally, it is assumed that hydrodynamic interactions are screened in polymer melts and that the properties of individual polymers in the  melt are well described by the Rouse model. \cite{dege:79,doi:86}  Screening is assumed to emerge by the excluded-volume interactions between the polymers. Conversely, analytical considerations show that in melts of phantom polymers, i.e., polymers without excluded-volume interactions, hydrodynamic interactions are unscreened. \cite{free:81}

Recent computer simulations and theoretical studies of unentangled  polymer melts including excluded-volume interactions raise considerable doubts on this simple pictures, since the studies show clear evidence of a long-time tail in the polymer velocity correlation function, indicative for unscreened hydrodynamic interactions. \cite{fara:11,fara:12.1}

In this article, we study the properties of a viscoelastic fluid by analytical calculations and simulations. Our goal is to characterize the properties of the viscoelastic fluid, which will ultimately be used to study  embedded objects. Analytically, we consider the linearized Navier-Stokes equations with a time-dependent relaxation modulus, i.e., an integro-differential equation for the velocity field. \cite{bird:87.1} The relaxation modulus follows from the Rouse model of polymer dynamics, \cite{doi:86} a special case of the generalized Maxwell model. \cite{bird:87.1}  In simulations, we employ the MPC approach, which has successfully been applied to study structural and dynamical properties of a wide range of polymeric systems. \cite{muss:05,ryde:06,fran:08,chel:12,niko:10,huan:10,huan:13,ripo:06,fedo:12,sing:14,wink:14.1,ghav:17} It correctly captures hydrodynamic interactions \cite{muss:05,huan:13} and can efficiently be parallelized on various platforms, especially on graphics processing units (GPUs). \cite{west:14,howa:18}

We analyze the fluid properties in terms of velocity autocorrelation functions. An analytical solution for the transverse velocity autocorrelation function is conveniently obtained in Fourier space, with respect to position, and in Laplace space, with respect to time. Inverse Laplace transformation yields a strongly time-dependent transverse autocorrelation function, which exhibits damped oscillations. Both, the damping and the oscillation frequencies depend on the relaxation times of the polymer and the wave vector. Independent of the polymer length, the (transverse) velocity autocorrelation function $C(t)$ exhibits a long-time tail on large length scales, with the time dependence $t^{-3/2}$ as is well established for Newtonian fluids. \cite{feld:05,alde:70,zwan:70,erns:71,haug:73,hinc:75,fara:12.1,huan:12} Hence, hydrodynamic correlations determine the dynamical properties of a melt of phantom polymers on large length scales.  This is reflected in the polymer center-of-mass diffusion coefficient, which exhibits the polymer length dependence according to the hydrodynamic Zimm model. \cite{doi:86}

The article is organized as follows. Section \ref{sec:fluid_model} presents the polymer model and  a description of the viscoelastic fluid in terms of a modified Navier-Stokes equation. Velocity autocorrelation functions of the fluid are introduced and their analytical solutions are presented in Sec.~\ref{section-analytical-solution-of-modified-Navier-Stokes}. The dynamics of the center-of-mass of an individual (tagged) polymer is discussed in Sec.~\ref{sec:com_dynamics}. Section~\ref{sec:mpc_implementation} describes the MPC implementation, and Sec.~\ref{sec:simulation_results} presents the simulation results and a comparison with theoretical predictions. Finally, the main results and aspects of our study  are summarized in Sec.~\ref{sec:conclusions}. The Appendices  \ref{app:Inverse-Laplace-Transform-of-VACFFTTLT} and \ref{app:inv_laplace}  describe  details of the calculation of  inverse Laplace transformations. Appendix C illustrates the derivation of the center-of-mass velocity autocorrelation function of a tagged polymer.

\section{Model of Viscoelastic Fluid} \label{sec:fluid_model}

\subsection{Polymer Dynamics} \label{sec:rouse_model}

We consider an ensemble of linear phantom polymers, each composed of $N$ monomers. The bonds between subsequent monomers are described by the harmonic Hamiltonian
\begin{align} \label{eq:hamiltonian}
H = \frac{K}{2} \sum_{i=1}^{N-1} \left( \bm r_{i+1} - \bm r_i \right)^2 .
\end{align}
In the stationary state, this leads to a Gaussian partition function capturing the conformational degrees of freedom of the polymer. \cite{wink:92}
The overdamped equation of motion for the position $\bm r_i(t)$ of monomer $i$, corresponding to the Rouse description of polymer physics, \cite{doi:86,harn:97} is then
\begin{align} \label{eq:langevin_rouse}
\dot{\bm r_i} = - \frac{1}{\gamma} \frac{\partial H}{\partial \bm r_i} + \frac{1}{\gamma} \bm \vGamma_i .
\end{align}
Here, $\dot{\bm r}_i(t)$ is the monomer  velocity at time $t$,
$k_B$ the Boltzmann constant, $T$ the temperature,  $\gamma$ the friction coefficient, and the $\bm \vGamma_i$ represent stationary, Markovian, and Gaussian random processes with zero mean and the second moments ($\alpha, \beta \in\{x,y,z\}$)
\begin{align}
\lla \vGamma_{i \alpha}(t) \vGamma_{j \beta}(t') \rra  = 2 k_BT \gamma \delta_{ij} \delta_{\alpha \beta} \delta(t-t') .
\end{align}
The coefficient $K$ in Eq.~\eqref{eq:hamiltonian} is related to the mean square bond length $l^2$ via $K=3k_BT/l^2$.

The solution of Eq.~\eqref{eq:langevin_rouse} is \cite{verd:66,kopf:97}
\begin{align}
\bm r_i(t) =  \sum_{n=0}^{N-1}  \bm \chi_n(t) b_n^{(i)} ,
\end{align}
 with the eigenfunctions
\begin{align}
 b_n^{(i)} = \sqrt{\frac{2}{N}}  \cos \left( \frac{n \pi}{N} \left[i-\frac{1}{2} \right] \right)
\end{align}
and
\begin{align}
 \bm \chi_0(t) = \sqrt{\frac{1}{2N}} \sum_{i=1}^N \bm r_i (t) =\sqrt{\frac{N}{2}} \bm r_{cm} .
\end{align}
The correlation functions of the mode amplitudes $\bm \chi_n$ are obtained as ($n, m \in \left[1,N-1\right]$)
\begin{align}
\lla \bm \chi_n(t) \cdot \bm \chi_m(t) \rra = \frac{l^2 \delta_{nm}}{4 \sin^2(n \pi/(2N))} e^{-t/\tau_n} ,
\end{align}
with the relaxation times
\begin{align} \label{eq:relax_time_disc}
\tau_n = \frac{\gamma l^2}{12 k_BT \sin^2( n\pi/2N)} .
\end{align}
In the continuum limit $N \to \infty$, $l \to 0$, such that $L=Nl$ remains constant, the well-know expression
\begin{align} \label{eq:relax_time_cont}
\tau_n = \frac{\gamma L^2}{3 \pi^2 k_BT n^2} = \frac{\tau_R}{n^2}
\end{align}
of the continuous Rouse model is obtained, with the Rouse relaxation time $\tau_R = \hat \gamma l L^2/3 \pi^2 k_BT$, the friction coefficient $\hat \gamma$ per length, and the bond length (Kuhn length) $l=2 l_p$, where $l_p$ is the persistence length. \cite{doi:86,harn:97}

The current formulation of the model, with $K=3 k_BT/l^2$, applies to equilibrium systems only, and cannot reproduce some nonequilibrium properties, such as shear thinning. To capture such effects, the stretching of polymer bonds by the external forces needs to be prevented. In case of simple shear, this is easily achieved by a shear-dependent coefficient $\mu(\dot \gamma)$ and the modified  force coefficient $K=3 \mu (\dot \gamma) k_B T/l^2$, where $\dot \gamma$ is the shear rate. The coefficient $\mu$ follows from the inextensibility constraint $\sum_{i=1}^{N-1} \langle (\bm r_{i+1} - \bm r_i)^2 \rangle = (N-1)l^2$.\cite{wink:99,wink:10} More general, the constraint $\langle (\bm r_{i+1} - \bm r_i)^2 \rangle = l^2$ for every bond can be applied with a corresponding number of Lagrangian multipliers. Even for dumbbells, shear thinning is obtained with this length constraint. \cite{kowa:13}

\subsection{Modified Navier-Stokes Equation}

The viscous properties of Newtonian fluids are described by the Navier-Stokes equations. \cite{land:59}
In the absence of external forces, the corresponding linearized equation for the fluid  momentum is
\cite{land:59}
\begin{align} \label{eq:navier_stokes}
 \varrho \frac{\partial \bm v(\bm r,t)}{\partial t} = - \nabla p + \eta \Delta \bm v(\bm r,t) ,
\end{align}
with  the fluid velocity   $\bm v(\bm r,t)$ and pressure $p(\bm r,t)$ fields  at the position $\bm r$ and time $t$, the fluid mass density $\varrho$, and the shear viscosity $\eta$.  We want to consider a viscoelastic fluid composed of the phantom polymers of Sec.~\ref{sec:rouse_model}. Viscoelasticity is incorporated in the Navier-Stokes equation by the (heuristic) extension
\cite{bird:87.1,ferr:80}
\begin{align} \label{eq:navier_stokes_visc}
 \varrho \frac{\partial \bm v(\bm r,t)}{\partial t} = - \nabla p + \int_0^t G(t-t') \Delta \bm v(\bm r,t') dt' .
\end{align}
Here, $G(t)$ is the shear relaxation modulus, which is independent of spatial coordinates
and vanishes in the asymptotic limit $(t-t^\prime) \to \infty$. \cite{bird:87.1} 

The relaxation modulus $G^p(t)$ for the phantom polymers of Sec.~\ref{sec:rouse_model} has been determined in Ref.~\onlinecite{doi:86} as ($t\ge 0$)
\begin{align}
G^p(t) = \varphi k_B T\sum_{n=1}^{N-1}  e^{-2t/\tau_n}  ,
\end{align}
where $\varphi$ is the number of polymers per volume.  The latter is related with the mass density $\varrho$ via
\begin{align} \label{eq:def_density}
\varphi = \frac{\varrho}{m N} = \frac{\phi}{N} ,
\end{align}
where $m$ is the  monomer mass, and $\phi$  the overall monomer concentration.
The complete fluid relaxation modulus $G(t)$ is, aside from the polymer-bond contribution, determined  by the ideal gas contribution of the individual monomers due to their thermal motion.  Hence, we use the relaxation modulus
\begin{align} \label{eq:modulus_disc}
G(t) = \eta \delta(t) + \varphi k_B T\sum_{n=1}^{N-1}  e^{-2t/\tau_n}  .
\end{align}
Then, the Navier-Stokes equation \eqref{eq:navier_stokes_visc} reduces to that of a  Newtonian fluid in case of a monomer solution ($N=1$).

The viscosity $\eta_f$ of the viscoelastic fluid follows from $G(t)$ via \cite{doi:86}
\begin{align}
\eta_f = \int_0^{\infty} G(t) dt ,
\end{align}
which yields, with Eq.~\eqref{eq:relax_time_cont},\cite{doi:86}
\begin{align} \label{eq:viscosity_f}
\eta_f = \eta + \frac{\varphi k_BT}{2} \sum_{n=1}^{N-1} \tau_n = \eta +\frac{\varphi \gamma l^2(N^2-1)}{36}.
\end{align}
With the density $\varphi$ of Eq.~\eqref{eq:def_density}, the fluid viscosity becomes
\begin{align} \label{eq:viscosity_f_c}
\eta_f =  \eta +\frac{\phi \gamma l^2}{36}\left(N- \frac{1}{N} \right).
\end{align}
For long polymers ($N\gg1$) and fixed $\phi$, $\eta_f$ is dominated by the bond contribution ($G^P$) and $\eta$ is negligible. Then, the fluid viscosity increases linearly with the degree of polymerization, $N$.

\section{Mesoscale Hydrodynamics: Multiparticle Collision Dynamics} \label{sec:mpc_implementation}

The MPC method for the simulation of polymer dynamics proceeds in two steps---streaming and collision. In the streaming step over a time interval $h$, where $h$ is denoted as collision time, Newton's equations of motion for the monomers,
\begin{align} \label{eq:polymer_mpc}
m \ddot{\bm r}_i = - \frac{\partial H}{\partial \bm r_i} ,
\end{align}
are solved by the velocity Verlet algorithm, \cite{Frenkel2002} with the Hamiltonian of Eq.~\eqref{eq:hamiltonian}. Since we consider phantom polymers, only bond forces contribute to the monomer dynamics. Other monomer-monomer interactions are implemented via MPC collisions.
Here, monomers are sorted into cubic cells of side length $a$, with the cells forming a complete tiling of the simulation volume, defining the collision environment.
We apply the Stochastic Rotation Dynamics (SRD) \cite{male:99,ihle:01,kapr:08} version of MPC, \cite{gomp:09} where the relative monomer velocities, with respect to the center-of-mass velocity of all monomers in a collision cell, are rotated around a randomly oriented axis by a fixed angle $\alpha$. This yields the new monomer velocities
\begin{align}
\bm v_i(t+h) = \bar{\bm v}_i(t+h) + \left(\mathrm{\bf R}(\alpha)-\mathrm{\bf E} \right) \left(\bar{\bm v}_i -\bar{\bm v}_{\textrm{cm}}^c \right) ,
\end{align}
where $\bar{\bm v}_i(t+h)$ is the monomer velocity after streaming, $\mathrm{\bf R}(\alpha)$ is the rotation matrix, $\mathrm{\bf E}$ the unit matrix,  and
\begin{align}
\bm v_{\textrm{cm}}^c(t)= \frac{1}{N_c}\sum_{j=1}^{N_c} \bm v_j(t)
\end{align}
is the center-of-mass velocity of the monomers in the cell of particle $i$;
$N_c$ is the total number of monomers in that particular cell. The random orientation of the rotation axis is chosen independently for every collision step and every collision cell.
Partitioning of the simulation volume into  collision cells implies violation of Galilean invariance.
To reestablish Galilean invariance,  a random shift of the collision lattice is performed at every collision step.\cite{ihle:01,gomp:09}
MPC conserves mass, momentum, and energy on the collision-cell level, which leads to correlations \cite{ripo:05,tuez:06} between the particles and long-range hydrodynamic interactions.\cite{huan:12}
To maintain a constant temperature, the Maxwell-Boltzmann-scaling (MBS) thermostat is applied
at every collision step and for every collision cell.\cite{Huang2010}

The simulations are performed with the hybrid program {\emph{OpenMPCD}}, \cite{OpenMPCD} a software suite implementing MPC-SRD \cite{male:99,Malevanets2000,gomp:09}
combined with molecular dynamics simulations (MD) (velocity Verlet algorithm \cite{Frenkel2002}).
Both, the MPC and the MD  part of the polymer dynamics---only phantom polymers with intramolecular bond interactions are considered---are executed in a massively parallel manner on a GPU (double precision). The program
exhibits excellent performance on graphical processing units (GPUs) \cite{west:14} supporting the CUDA%
\cite{CUDA-C-Programming-Guide-7.5} programming framework, such as NVIDIA Tesla accelerators.

Dimensionless units are introduced by scaling length by the cell size $a$, energy by $k_BT$, and time by $\sqrt{m a^2/k_BT}$. This corresponds to the choice $a=k_BT=m=1$. We choose the collision time $h=0.1\sqrt{m a^2/k_BT}$, the rotation angle $\alpha = 2.27 \mathrm{rad} \approx 130^{\circ}$, and the mean number of monomers in a collision cell $\lla N_c \rra =10$. The latter is equivalent with the mean fluid density $\varrho = 10  m/a^3$. A MD time step of $\Delta t = 0.02 \sqrt{m a^2/k_BT}$, smaller than the collision time step, is used in order to resolve the polymer dynamics adequately.
Three-dimensional periodic systems are considered with a cubic simulation box of side length $L_S=30 a$, if not indicated otherwise, corresponding to a total number of $N_{tot} = 2.7 \times 10^5$ monomers/MPC particles. Simulations of a monomer fluid, i.e., a bare MPC fluid,  yield the viscosity $\eta/\sqrt{mk_BT/a^4} =8.7$. \cite{huan:15} In the following, the units $a$, $k_BT$, and $m$ will be dropped, i.e., are set to unity. For the results presented in Sec. VI, between $2 \times 10^7$ and $1 \times 10^8$ MPC steps have been performed, typically approximately $4 \times 10^7$.

Simulations are initialized by placing the first monomer of every polymer at a random point in the
simulation volume sampled from a uniform distribution. Subsequent bound monomers are placed randomly by choosing a randomly oriented  unit bond vector. Initial velocities of each monomer are assigned independently with Cartesian components taken from a standard normal distribution.

\section{Velocity Correlation Function of Viscoelastic Fluid}
\label{section-analytical-solution-of-modified-Navier-Stokes}

The linear equation \eqref{eq:navier_stokes_visc} can be solved using Fourier and Laplace transforms.  Spatial Fourier transformation (denoted by a tilde) of the velocity,
\begin{align} \label{eq:ft_inf}
\tilde{\bm v}(\bm k,t) = \int \bm v(\bm r,t) e^{-i \bm k \cdot \bm r} d^3r ,
\end{align}
where $\tilde{\bm v}(\bm k,t)$ denotes the transformed velocity, yields
\begin{align} \label{eq:fourier-transformed-Navier-Stokes}
\varrho \frac{\partial \tilde{\bm v}(\bm k,t)}{\partial t} =  -i \bm k {\tilde p}(\bm k,t) - {\bm k}^2 \int_0^t G(t-t') \tilde{\bm v}(\bm k,t') dt' .
\end{align}
By multiplying this equation with $\tilde{\bm v}(-\bm k,0)$, we obtain
\begin{eq} \label{eq:VACFFTT-equation-in-Fourier-space}
\varrho \frac{\partial \tilde{C}^T(\bm k,t)}{\partial t} = - {\bm k}^2 \int_0^t G(t-t')\tilde{C}^T(\bm k,t') dt'
\end{eq}
for the transverse velocity autocorrelation function
\begin{eq} \label{eq:def_velcor}
{\tilde C}^T(\bm k,t) = \lla \tilde{\bm v}^T(\bm k,t) \cdot \tilde{\bm v}^T(-\bm k,0) \rra,
\end{eq}
where the brackets denote statistical averaging. The transverse component $\tilde{\bm v}^T(\bm k,t)$ is the component of the Fourier-space velocity $\tilde{\bm v}(\bm k,t) = \tilde{\bm v}^L(\bm k,t) + \tilde{\bm v}^T(\bm k,t)$ that is perpendicular to the Fourier vector $\bm k$, i.e., $\bm k \cdot \tilde{\bm v}^T(\bm k,t) = 0$. Laplace transformation (denoted by a circumflex) with respect to time,
\begin{align} \label{eq:laplace}
\hat{C}^T(\bm k,s) = \int_0^{\infty} \tilde{C}^T(\bm k,t) e^{-st} dt ,
\end{align}
yields
\begin{eq} \label{eq:VACFFTTLT}
\hat{C}^T(\bm k,s) = \frac{\varrho \tilde{C}^T(\bm k,0)}{\varrho s + \bm k^2 \hat G(s)} .
\end{eq}
We assume that the system is in thermal equilibrium at $t=0$, hence, \cite{huan:12}
\begin{align} \label{eq:corr_equil}
\tilde C^T(\bm k, 0) \equiv \tilde C^T(0) = \frac{2 k_BT}{\varrho} .
\end{align}
The Laplace transform of $G(t)$, Eq.~\eqref{eq:modulus_disc}, is \cite{Dyke2014,
ober:12} 
\begin{eq}
\hat G(s) = \eta + \varphi k_BT \sum_{n=1}^{N-1} \frac{1}{s + 2/\tau_n} ,
\end{eq}
and we thus obtain
\begin{eq}\label{Rouse-VACFFTTLT}
\hat{C}^T(\bm k,s) = \displaystyle  \frac{\displaystyle \varrho \tilde{C}^T(0)}{\displaystyle  \varrho s + \bm k^2 \left(\eta + \varphi k_BT \sum_{n=1}^{N-1} \frac{1}{s + 2/\tau_n}  \right) } .
\end{eq}
The explicit expression $\tilde{C}^T(\bm k,t)$ for the inverse Laplace transform
of this function is presented in
\eqnref{VACFFTT-general-solution} of Appendix \ref{app:Inverse-Laplace-Transform-of-VACFFTTLT}.%

The velocity correlation function $C(t) = \langle \bm v(\bm r, t) \cdot \bm v(\bm r, 0) \rangle$ follows by inverse Fourier and Laplace transformation. To eliminate the spatial dependence,  we average the correlation function with the  distribution function for $\bm r(t)$. \cite{alde:70,erns:71,huan:12} Adopting the Lagrangian description of the fluid, where a fluid element is followed as it moves through space and time, we obtain in general
\begin{align} \nonumber
C(t) & = \frac{1}{(2\pi)^3} \int \tilde C(\bm k,t) \lla e^{i \bm k \cdot( \bm r(t) - \bm r(0))} \rra  \ d^3k  \\ \label{eq:vel_corr_time}
& = \frac{1}{(2\pi)^3} \int \tilde C(\bm k,t) e^{- \bm k^2 \langle ( \bm r(t) - \bm r(0))^2 \rangle /6}   \ d^3k ,
\end{align}
due to the Gaussian nature of  the displacement distribution function. \cite{doi:86}  The mean square displacement (MSD), averaged over all monomers, is \cite{doi:86}
\begin{align}
\lla \Delta \bm r(t)^2 \rra = \lla ( \bm r(t) - \bm r(0))^2 \rra = 6 D_{\textrm{cm}} t + \lla \Delta \bm r(t)^2_m \rra ,
\end{align}
where $D_{\textrm{cm}} = k_BT/\gamma N$ is the center-of-mass diffusion coefficient, \cite{doi:86} and with the results of  Sec.~\ref{sec:rouse_model}, we obtain the average monomer MSD in the polymer center-of-mass reference frame:
\begin{align}
\lla \Delta \bm r(t)^2_m \rra = \frac{l^2}{2N} \sum_{n=1}^{N-1} \frac{1}{\sin^2 (n\pi/(2 N))} \left( 1- e^{- t/\tau_n} \right) .
\end{align}
Examples of the correlation function $G(t)$ for various polymer lengths are discussed in the following.

\subsection{Newtonian Fluid ($N=1$)} \label{sec:corr_n1}

A Newtonian fluid is recovered for $N=1$, and correspondingly $G(t) = \eta \delta(t)$.
The inverse Laplace transformation of
\begin{align}
\hat{C}^T(\bm k,s) = \frac{\tilde{C}^T(0)}{s + \bm k^2 \nu} ,
\end{align}
with the kinematic viscosity $\nu=\eta/\varrho$, yields
the time-dependent velocity-correlation function in Fourier space, \cite{ober:12}
\begin{align} \label{VACFFTT-Newtonian-fluid}
\tilde{C}^T(\bm k,t) =  \tilde{C}^T(0) e^{-\nu \bm k^2 t} ,
\end{align}
in agreement with previous studies. \cite{huan:12} With Eq.~\eqref{eq:corr_equil}, the correlation function $C(t) =C^L(t)+ C^T(t)$ of Eq.~\eqref{eq:vel_corr_time} becomes, in the long-time limit, \cite{huan:12,feld:05,alde:70,zwan:70,erns:71}
\begin{align} \label{eq:corr_mpc}
C(t)\approx C^T(t) = \frac{k_BT}{4 \varrho} \frac{1}{(\pi [\nu + D_{\textrm{cm}}]t)^{3/2}} ,
\end{align}
since the contribution of the longitudinal velocity correlation, $C^L(t)$, decays exponentially. \cite{huan:12,huan:13}

\subsection{Dumbbell fluid ($N=2$)} \label{sec:dumbbell}

Polymer-like aspects are already captured by a dumbbell (dimer)---i.e., two bound monomers---at least as long as the longest relaxation time of a  polymer dominates its internal dynamics. Here, \eqnref{VACFFTT-general-solution} assumes the form
\begin{align}  \label{eq:VACFFTT-dimer-simple-root}
\tilde{C}^T(\bm k,t) = & \ \frac{\tilde{C}^T(\bm k,0)}{\omega} e^{-\zeta t} \\ \nonumber & \times \left(\left[\frac{2}{\tau_1} - \zeta \right] \sin(\omega t) + \omega \cos(\omega t) \right) ,
\end{align}
with the abbreviations
\begin{align} \label{eq:damping_db}
\zeta =  & \ \frac{1}{2} \left( \frac{2}{\tau_1} + \bm k^2 \nu \right) ,  \\ \label{eq:omega_db}
\omega = & \ \frac{1}{\tau_1} \sqrt{2\bm k^2 \tau_1 \left(\nu_f -\nu \right) - \frac{1}{8} \left( \bm k^2 \nu \tau_1 -2 \right)^2} ,
\end{align}
and the kinematic viscosity $\nu_f=\eta_f/\varrho=\nu+\varphi k_BT\tau_1/(2 \varrho) > \nu$.
The correlation function~\eqref{eq:VACFFTT-dimer-simple-root} exhibits exponentially damped oscillations, where both the frequency, $\omega$, and the damping, $\zeta$, depend on the relaxation time $\tau_1$.

Evidently,  the radicand in Eq.~\eqref{eq:omega_db} is always negative for $\nu = \nu_f$. More general, in case of a negative radicand, the substitution $\omega = i \lambda$, with
\begin{align} \label{eq:lambda_db}
\lambda =  \frac{1}{\tau_1} \sqrt{-2\bm k^2 \tau_1 \left(\nu_f -\nu \right) + \frac{1}{8} \left( \bm k^2 \nu \tau_1 -2 \right)^2} ,
\end{align}
yields the  correlation function
\begin{align}  \label{eq:VACFFTT-dimer-damped}
\tilde{C}^T(\bm k,t) = & \ \frac{\tilde{C}^T(\bm k,0)}{\lambda} e^{-\zeta t} \\ \nonumber & \times \left(\left[\frac{2}{\tau_1} - \zeta \right] \sinh(\lambda t) + \lambda \cosh(\lambda t) \right) .
\end{align}
Since $\zeta > \lambda$, we obtain a non-oscillating correlation function. Equation~\eqref{eq:omega_db}, or Eq.~\eqref{eq:lambda_db}, clearly reveal a qualitative different dynamical behavior due to polymer elasticity (viscoelasticity). An oscillatory correlation function function appears for $\nu_f> \nu$ only. There are two obvious limits with only exponentially decaying correlation functions, namely $|\bm k| \to 0$ and $|\bm k| \to \infty$, which correspond to large and small scales, respectively.

In the limit $\lambda t \gg 1$, Eq.~\eqref{eq:VACFFTT-dimer-damped} becomes
\begin{align}  \label{eq:VACFFTT-dimer-limit_t}
\tilde{C}^T(\bm k,t) =  \tilde{C}^T(\bm k,0) \left( \frac{1}{\lambda \tau_1} - \frac{\zeta}{2 \lambda} + \frac{1}{2}  \right) e^{-(\zeta - \lambda) t} .
\end{align}
For $|\bm k| \to 0$, the difference in the exponent reduces to $\zeta - \lambda = \bm k^2 ( \nu + \varphi k_BT \tau_1/2 \varrho) + O(\bm k^4)$ (Eq.~\eqref{eq:viscosity_f}), and the correlation function decays exponentially with the total kinematic viscosity, $\nu_f$,
\begin{align}  \label{eq:VACFFTT-dimer-limit_t_exp}
\tilde{C}^T(\bm k,t) =  \tilde{C}^T(\bm k,0)e^{- \nu_f \bm k^2 t} .
\end{align}
Then, Fourier transformation w.r.t.\ $\bm k$ of Eq.~\eqref{eq:VACFFTT-dimer-limit_t_exp} yields a long-time tail $C^T(t) \sim (\nu_f t)^{-3/2}$ on large length and long time scales.
Conversely, on
small length scales $|\bm k| \to \infty$, the exponent becomes $\zeta - \lambda = 2 (1 + \nu_f/ \nu )/ \tau_1 + O(1/ \bm k^2)$, and the decay of the correlation function depends only weakly on the wave vector. In both cases, the decline of
$\tilde{C}^T(\bm k,t)$ is determined by the properties of the dumbbell rather than the individual monomers.

\begin{figure}[t]
\begin{center}
\includegraphics[width=\columnwidth]{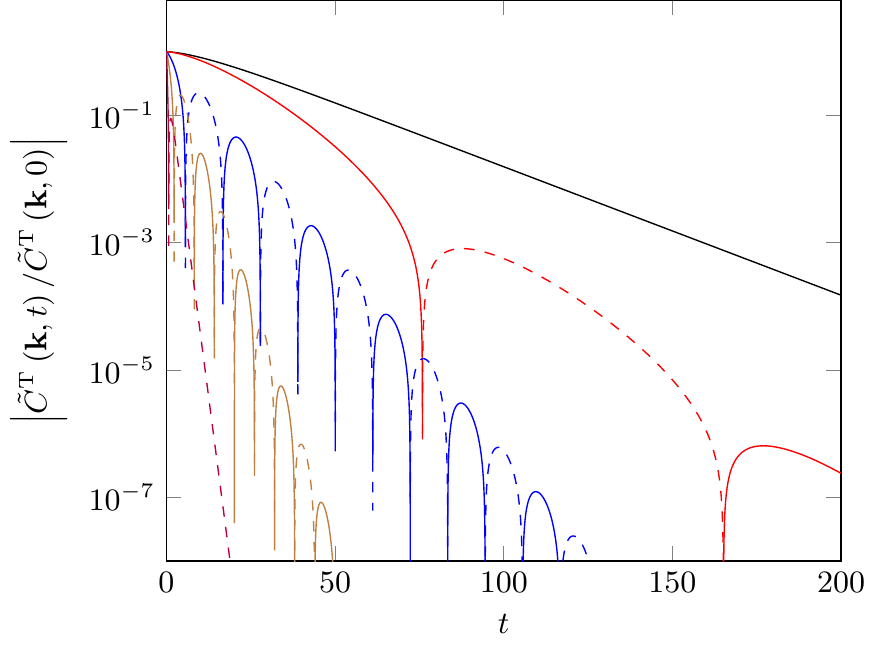}
\end{center}
\caption{Transverse velocity autocorrelation function $\tilde C^T(\bm k,t)/\tilde C^T(\bm k,0)$ of dumbbells (Eq.~\eqref{eq:VACFFTT-dimer-simple-root} or \eqref{eq:VACFFTT-dimer-damped}) as a function of time for the wave vectors $|\bm k| = 0.09$, $0.11$, $0.4$, $0.8$, and $2$ (top to bottom). The other parameters are
	$\eta = 8.7$,
	$\varrho = 10$,
	$\varphi k_B T = 5$, and $\tau_1 = 13.4$.  Due to viscoelasticity, the correlation function oscillates for certain wave vectors assuming positive (solid) and negative (dashed) values.
} \label{fig:corr_dumbbell}
\end{figure}

Figure~\ref{fig:corr_dumbbell} provides examples of the correlation function $\tilde C^T(\bm k,t)$ for various wave vectors and a specific set of parameters (see figure caption).
Note that the oscillating correlation functions assume positive and negative values.  The correlation functions for the smallest ($k=0.09$) and the largest ($k=2$)  displayed $\bm k$ vectors decay exponentially according to Eq.~\eqref{eq:VACFFTT-dimer-limit_t}, whereas those for in-between $\bm k$ values exhibit exponentially damped oscillations, Eq.~\eqref{eq:VACFFTT-dimer-simple-root}. In the latter case, the oscillation frequency increases with increasing wave number. The decay rates in the limit $|\bm k| \to 0$ and $|\bm k| \to \infty$ agree with the values discussed above.

\pgfmathsetmacro{\myVACFFTZero}{1.0}
\pgfmathsetmacro{\myEta}{8.705}
\pgfmathsetmacro{\myRho}{10.0}
\pgfmathsetmacro{\myBaseKSquared}{(2 * pi / 30.0)^2}
\pgfmathsetmacro{\myFirstTau}{13.43}
\pgfmathsetmacro{\myPhiKT}{\myRho / 2.0}

\def\myParameterList{1/13.43, 2/13.43, 1/20.0, 1/40.0} 

\subsection{Continuous Polymer ($N \to \infty$)} \label{sec:corr_cont}

In the case of continuous polymers, $N\to\infty$, the relaxation modulus~\eqref{eq:modulus_disc} becomes
\begin{align} \label{eq:relax_mod_cont}
G(t) = \eta \delta(t) +  \varphi k_BT \sum_{n=1}^{\infty} e^{-2 tn^2/\tau_R} .
\end{align}
Two limiting case can be distinguished
\begin{itemize}[leftmargin=*]
\item[(i)] $t/\tau_R \gg1$ ---
The sum in Eq.~\eqref{eq:relax_mod_cont} is dominated by the mode $n=1$,
corresponding to the dumbbell considered in Sec.~\ref{sec:dumbbell} with the relaxation time $\tau_1 = \tau_R$. Again, on large length scales, the velocity correlation function decays exponentially with the viscosity $\eta_f = \eta + \varphi k_BT \tau_R/2$, where $\eta$ can be neglected for long polymers due to the large Rouse time.
\item[(ii)] $t/\tau_R \ll 1$ --- The sum of modes in Eq.~\eqref{eq:relax_mod_cont} can be replaced by an integral
over $n$, \cite{doi:86,wink:06}
and we straightforwardly obtain
\begin{align}
G(t) =   \eta \delta(t)  + \frac{\varphi k_BT}{2} \sqrt{\frac{\pi \tau_R}{2 t}} .
\end{align}
Laplace transformation yields
\begin{align}
\hat G(s) = \eta +  \frac{\varphi k_BT \pi }{2} \sqrt{\frac{\tau_R}{2 s}} ,
\end{align}
and the velocity correlation function becomes
\begin{align} \label{eq:corr_cont_poly_s}
\hat{C}^T(\bm k,s) = \frac{\displaystyle \varrho \tilde{C}^T(\bm k,0)}{\displaystyle  \varrho s + \bm k^2 \left(\eta + \frac{\varphi k_BT \pi }{2} \sqrt{\frac{\tau_R}{2 s}} \right)} .
\end{align}
Inverse Laplace transformation (cf. Appendix~\ref{app:inv_laplace}) yields, neglecting $\eta$,
\begin{align} \label{eq:corr_scal}
\tilde C^T(\bm k, t) =  {\tilde C}^T(0) f([\bm k^2 \varphi k_BT \pi \sqrt{\tau_R} / (2 \sqrt{2} \varrho)]^{2/3} t) ,
\end{align}
with the function $f(x)$ specified in Eq.~\eqref{app:int_laplace}.
Hence, $\tilde C^T(\bm k, t)$ scales with $(k^{4/3}t)$ as already pointed out in Ref.~\onlinecite{fara:12.1}.

In the asymptotic limit of a large argument of $f$ in Eq.~\eqref{eq:corr_scal}, we find
\begin{align} \label{eq:corr_approx_time}
\tilde C^T(\bm k, t) =  - \frac{ \sqrt{2} \varrho \, {\tilde C}^T(\bm k,0)}{\varphi k_BT \pi \sqrt{\pi\tau_R}} \frac{1}{\bm k^2 t^{3/2}} .
\end{align}
Note that the correlation function is negative.
The correlation function  exhibits a long-time-tail-type time dependence  $t^{-3/2}$, which is  very different from the exponential function of Eq.~\eqref{VACFFTT-Newtonian-fluid} for individual monomers. Polymer elasticity complete changes the time and wave-vector dependence of the correlation function. However, as shown in Sec.~\ref{sec:corr_cm_cont}, this does not affect the long-time-type decay of the correlation function in real space.

Inserting $\tilde C^T(\bm k,0)$ of Eq.~\eqref{eq:corr_equil} and the polymer concentration of Eq.~\eqref{eq:def_density} into Eq.~\eqref{eq:corr_approx_time}, we find
\begin{align} \label{eq:corr_approx_time_l}
\tilde C^T(\bm k, t) =  - \frac{2 \sqrt{2} }{\pi^{3/2} \phi l \sqrt{\tau_l}} \frac{1}{\bm k^2 t^{3/2}} ,
\end{align}
with the abbreviation $\tau_l = \hat \gamma l/(3\pi^2 k_BT)$.
Thus, the fluid correlation function $\tilde C(\bm k,t)$ is independent of polymer length in the time interval $\tau_l \ll t \ll \tau_R$ for long polymers ($L/l \gg 1$);
it depends on the overall monomer density only.
\end{itemize}

\subsection{Asymptotic Behavior for $t \to \infty$} \label{sec:corr_asymp}

The asymptotic time dependence of the correlation function $C^T(t)$ for $t \to \infty$ follows from Eq.~\eqref{Rouse-VACFFTTLT} in the limit $s \to 0$. Neglecting $s$ in the sum over modes in Eq.~\eqref{Rouse-VACFFTTLT}, the correlation function reduces to $\hat C^T(\bm k,s) =\tilde C^T(0)/(s + \bm k^2 \nu_f)$, with the total kinematic viscosity $\nu_f = \eta_f/\varrho$. Then, Fourier and Laplace transformations yield
\begin{align} \label{eq:corr_asymp_t}
C^T(t) = \frac{\tilde C^T(0)}{8  \left( \pi \nu_f t \right)^{3/2}}
\end{align}
independent of polymer length.  This result is consistent with the limiting cases discussed in  Secs.~\ref{sec:corr_n1}, \ref{sec:corr_cont}, and corresponds to the long-time tail of simple fluids.\cite{huan:12,feld:05,alde:70,zwan:70,erns:71} Hence, on large length scales and for long times, the polymer melt of phantom chains  exhibits fluid-like behavior similar to simple fluids, but with the total viscosity $\eta_f$ determined by polymer elasticity.

\subsection{Oseen Tensor-Type Behavior}

Integration of the correlation function ${\tilde C}^T(\bm k,t)$
over time yields
\begin{align}
T(k) = \int_0^{\infty} {\tilde C}^T(\bm k,t) dt =  {\hat C}^T(\bm k,0) = \frac{\varrho \tilde C^T(0)}{\bm k^2 \eta_f} ,
\end{align}
with the use of the definition of the Laplace transform \eqref{eq:laplace} and Eq.~\eqref{Rouse-VACFFTTLT}.
Hence, $T(k) \sim 1/\bm k^2$,  similar to the Oseen tensor of a Newtonian fluid, \cite{doi:86,dhon:96,huan:12} but with the viscosity of the polymeric fluid.
Fourier transformation with respect to $\bm k$ leads then to a long-range interaction $\sim 1/|\bm r|$ in three-dimensional space. As a consequence, the considered viscoelastic phantom polymer melt exhibits the properties of Newtonian fluids in terms of long-range fluid interactions.

\section{Center-of-Mass Dynamics of  Individual Polymers} \label{sec:com_dynamics}

The equations of motion \eqref{eq:langevin_rouse} describe the dynamics of an isolated polymer exposed to thermal noise. We now consider an individual polymer embedded in other, identical polymers,  accounting for  the environment by including the convective transport velocity following from Eq.~\eqref{eq:navier_stokes_visc}. Hence, we set
\begin{align} \label{eq:convection}
\dot{\bm r}_i (t) = \bm v(\bm r_i,t) -  \frac{1}{\gamma} \frac{\partial H}{\partial \bm r_i} + \frac{1}{\gamma} {\bm \varGamma}_i(t),
\end{align}
with the fluid velocity  $\bm v(\bm r_i,t)$ at the location of monomer $i$. Considering the center-of-mass motion only, we find
\begin{align}
\dot{\bm r}_{\textrm{cm}} (t) =  \frac{1}{N} \sum_{i=1}^N \bm v(\bm r_i,t) + \frac{1}{ \gamma N} \sum_{i=1}^N{\bm \varGamma}_i(t) ,
\end{align}
for the position $\bm r_{\textrm{cm}}$ and velocity $\dot{\bm r}_{\textrm{cm}}$
of the center of mass of a particular polymer.

\subsection{Center-of-Mass Velocity Correlation Function} \label{sec:com_corr}

The center-of-mass velocity autocorrelation function is given by ($t>0$)
\begin{align} \nonumber \label{eq:corr_cm_def}
C_{\textrm{cm}}(t) & \ =  \lla \bm \dot{\bm r}_{\textrm{cm}} (t) \cdot \dot{\bm r}_{\textrm{cm}} (0) \rra \\
& \ = \frac{1}{N^2} \sum_{i=1}^N \sum_{j=1}^N
\lla \bm v(\bm r_i,t) \cdot \bm v(\bm r_j,0) \rra .
\end{align}
Focusing on the transverse velocity correlation function, in Fourier representation, we obtain \cite{fara:12.1,huan:13} (cf. App.~\ref{app:com_corr} for details)
\begin{align} \label{eq:com_corr}
C_{\textrm{cm}}^T(t) = \frac{1}{(2\pi)^3N} \int S(\bm k,t) {\tilde C}^T(\bm k,t) d^3 k,
\end{align}
with   ${\tilde C}^T(\bm k,t)$ presented in Sec.~\ref{section-analytical-solution-of-modified-Navier-Stokes} and the dynamic structure factor  \cite{doi:86,harn:96}
\begin{align} \nonumber \label{eq:dyn_struct_gen}
S(\bm k,t) = &~\frac{1}{N} \sum_{i=1}^N \sum_{j=1}^N \lla e^{i \bm k \cdot (\bm r_i(t) - \bm r_j(0))} \rra \\
=  &~\frac{1}{N} \sum_{i=1}^N \sum_{j=1}^N \exp\left( -\bm k^2 \langle ( \bm r_i(t) -\bm r_j(0))^2\rangle/6\right)
\end{align}
of the polymer.
Note that the solution of the polymer dynamics of Sec.~\ref{sec:rouse_model} yields a Gaussian distribution of the monomer-monomer separation $\bm r_i - \bm r_j$, with $\langle (\bm r_i - \bm r_j)^2 \rangle = |i-j|l^2$, and the mean square displacement \cite{wink:97}
\begin{align} \label{eq:msd_gen}
 \lla  ( \bm r_i(t) -\bm r_j(0))^2 \rra &= |i-j|l^2 + 6 D_{cm} t \\ \nonumber &  + \frac{6 k_BT}{\gamma}  \sum_{n=1}^{N-1} \tau_n b_n^{(i)} b_n^{(j)} \left(1-e^{-t/\tau_n} \right) .
\end{align}
Strictly speaking, the mean square displacement has to be obtained from the solution of Eq.~\eqref{eq:convection}, which includes the convective flow field induced by neighboring polymers. However, on large length scales and for long polymers, the most significant contribution comes from small $\bm k$ values. Hence, the time dependence of the dynamic structure factor can be neglected and the static structure factor, $S(\bm k,0)$, can be used, i.e., the relevant properties of the tracer polymer are captured by its equilibrium structure. In general, the term $D_{cm}t$ can be neglected for long polymers, because the kinematic viscosity is typically much larger than $D_{cm}$ (cf. Eq.~\eqref{eq:corr_mpc}). Our calculations confirm that these approximations apply, and that $C_{cm}(t)$ is essentially identical when using either $S(\bm k,t)$ or $S(\bm k,0)$, even
for dumbbells. Consequently, the time dependence of $C_{cm}(t)$ is completely determined by the correlation function,
$\tilde C(\bm k,t)$, of the viscoelastic fluid.

\subsubsection{Dumbbell fluid ($N=2$)}

For a dumbbell, the dynamics structure factor \eqref{eq:dyn_struct_gen} reads
\begin{align} \nonumber \label{eq:dyn_strcut_dumb}
S(\bm k,t) = & e^{- D_{cm} \bm k^2  t} \left[\exp\left(- \frac{\bm k^2 l^2}{12} \left(1-e^{-t/\tau_1}  \right) \right) \right. \\  & + \left.  \exp\left(\frac{\bm k^2 l^2}{12} \left(1-e^{-t/\tau_1} \right) \right)  e^{-\bm k^2 l^2/6} \right].
\end{align}
The correlation function $C_{\textrm{cm}}^T(t)$ is then obtained by evaluation of  Eq.~\eqref{eq:com_corr} with the correlation function  ${\tilde C}^T(\bm k,t)$ of Sec.~\ref{sec:dumbbell}.

For a dumbbell, the contribution of the convective velocity in Eq.~\eqref{eq:convection} to dynamic properties, e.g., an effective relaxation time, is negligible as shown in Ref.~\onlinecite{kowa:13}, and the relaxation time of the Rouse model can be used. As pointed out above, the numerical evaluation of Eq.~\eqref{eq:com_corr} yields essentially the same correlation function when using either $S(\bm t,t)$  or $S(\bm k,0)$.

\subsubsection{Continuous Polymer $(N \to \infty)$} \label{sec:corr_cm_cont}

In the limit of a continuous polymer, integration of Eq.~\eqref{eq:com_corr}  with the correlation function \eqref{eq:corr_approx_time_l} yields
\begin{align} \label{eq:corr_cm_cont}
C_{\textrm{cm}}^T(t) = - \frac{8 }{\sqrt{3 l^3} \pi^3 \phi \sqrt{\tau_l}} \frac{1}{t^{3/2} \sqrt{L}}
\end{align}
for the time interval $\tau_l \ll t \ll \tau_R$. We take only the static structure factor into account in deriving Eq.~\eqref{eq:corr_cm_cont}, i.e., we set $t=0$ in Eq.~\eqref{eq:msd_gen}.  As our numerical studies show, a more precise account of $S(\bm k,t)$ changes the very short-time behavior of the correlation function, but does not affect the longer-time decay, which is of primary interest here.

Evidently, the correlation function \eqref{eq:corr_cm_cont} exhibits a power-law decay $t^{-3/2}$ reminiscent to the long-time tail of hydrodynamics.\cite{huan:12,feld:05,alde:70,zwan:70,erns:71} However, the asymptotic time regime for $t \to \infty$, corresponding to the long-time-tail hydrodynamics of simple fluids, is described by Eq.~\eqref{eq:corr_asymp_t}. The correlation function \eqref{eq:corr_cm_cont} emerges from the polymer character of the fluid, with its nearly continuous mode spectrum. The coupling of internal polymer dynamics leads to fluid-like large-scale and long-time correlations. Similar dependencies on time and polymer length, $1/\sqrt{L}$, have been obtained in Ref.~\onlinecite{fara:12.1}.

\subsection{Center-of-Mass Diffusion}

The center-of-mass diffusion coefficient follows from the  center-of-mass correlation function via the relation
\begin{align}
D = \frac{1}{3} \int_0^{\infty} C_{\textrm{cm}}^T(t) dt = \frac{1}{3 (2\pi)^3 N} \int S(\bm k)\hat C^T(\bm k,0) d^3k ,
\end{align}
because the integral over the longitudinal contribution of the correlation function vanishes.  \cite{huan:12,huan:13}
Evaluation of the integral with Eqs.~\eqref{eq:corr_equil} and \eqref{Rouse-VACFFTTLT} yields
\begin{align}
D = \frac{8 k_BT}{3\sqrt{6 \pi^3} \eta_f} \frac{1}{\sqrt{lL}}
\end{align}
for a continuous polymer. This is the diffusion coefficient of a polymer in a solution of viscosity $\eta_f$ (Zimm model).\cite{doi:86}  Thus, our phantom-polymer melt yields the same polymer length dependence, i.e.,  $1/\sqrt{lL}$, as a polymer in solution. This emphasizes that hydrodynamic interactions are fully developed and determine the diffusive behavior.

\section{Simulation Results, Comparison with Theory} \label{sec:simulation_results}

\subsection{Correlation Function $\tilde C^T(\bm k,t)$}

In simulations, periodic boundary conditions are applied, and the monomer velocities in Fourier space are calculated as
\begin{align}
\tilde {\bm v}(\bm k,t)  = \frac{1}{N_{\textrm{tot}}} \sum_{i=1}^{N_{\textrm{tot}}} \bm v_i(t) e^{-i \bm k \cdot \bm r_i(t)} .
\end{align}
Due to the periodic boundary conditions, the Cartesian components $k_{\alpha}$ of the wave vector
$\bm k = (k_x, k_y, k_z)^T$ assume the values $k_{\alpha}=2 \pi n_{\alpha}/L$, with
$n_{\alpha} \in \mathbb{Z}$, $\alpha \in \{x,y,z\}$, and $N_{\textrm{tot}}=NN_p$ the total number of monomers.
Note that only $k$-values with $|\bm k| \ne 0$ are allowed.
Here, the Fourier transformation [Eq.~\eqref{eq:ft_inf}] of the viscoelastic continuum is adjusted to periodic boundary conditions as described in Ref.~\onlinecite{huan:12}. In agreement with the results of Ref.~\onlinecite{huan:12}, the transverse velocity correlation function of the bare MPC fluid (monomers) decays exponentially according to Eq.~\eqref{VACFFTT-Newtonian-fluid} with the kinematic viscosity $\nu=0.87$.

\begin{figure}[t!]
\begin{center}
\includegraphics[width=\columnwidth]{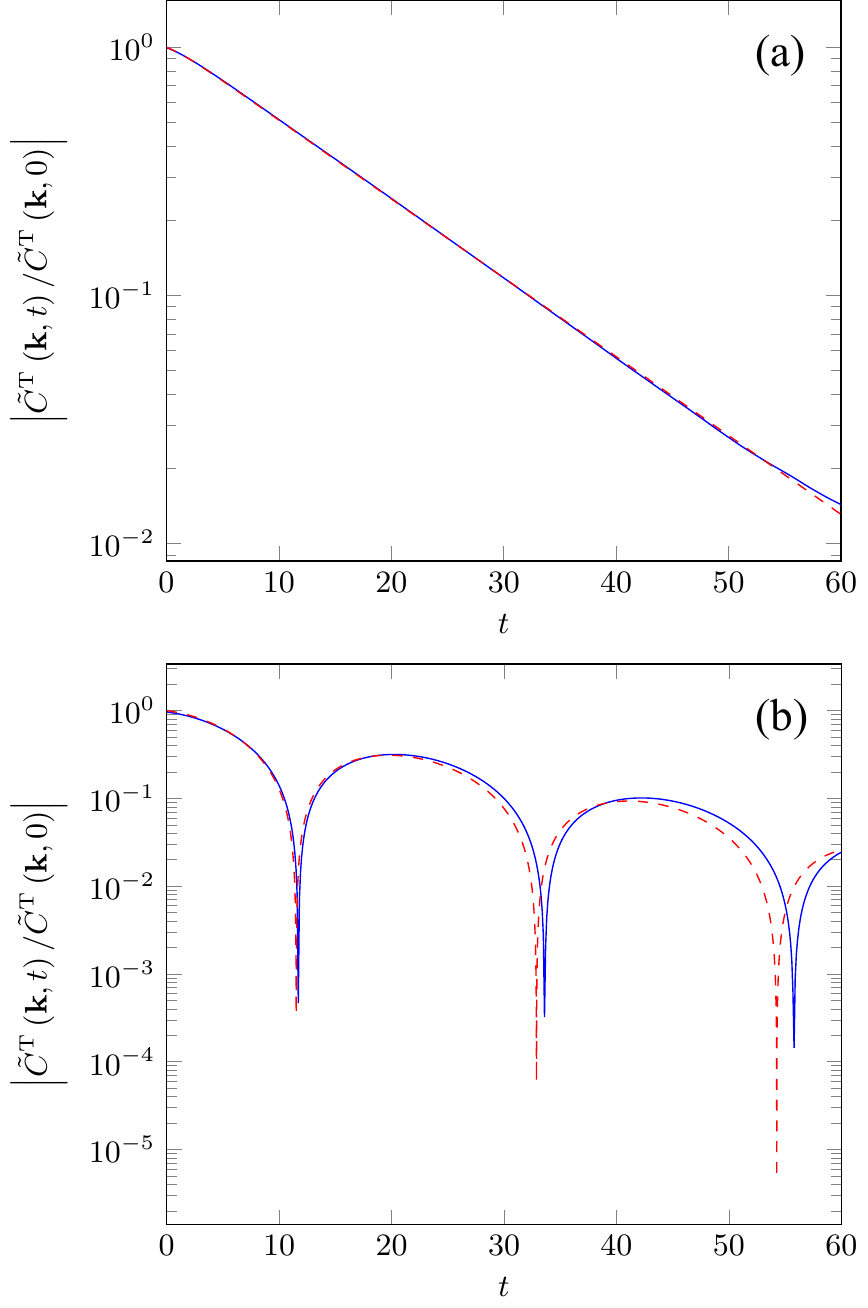}
\end{center}
\caption{Transverse velocity autocorrelation functions of dumbbells, i.e., $N=2$,
from simulations (solid) and analytical theory (dashed), Eqs.~\eqref{eq:VACFFTT-dimer-simple-root}, \eqref{eq:VACFFTT-dimer-damped},
as a functions of time $t$
for the bond length (a) $l=1$ ($K=3$) and (b) $l=\sqrt{10}$ ($K=0.3$).
The wave vector is in all cases $\bm k = (2\pi/30,0,0)^T$. As in Fig.~\ref{fig:corr_dumbbell}, the correlation function in (b) assumes negative values (even loops). The theoretical curves are fitted to the simulation data.
} \label{fig:corr_dumbbell_sim_theory}
\end{figure}

\subsubsection{Dumbbell Fluid ($N=2$)}

Results for the transverse velocity autocorrelation function of dumbbells of various bond lengths are displayed in Fig.~\ref{fig:corr_dumbbell_sim_theory}.  We like to mention that accurate simulation data for long times are rather demanding in terms of simulation time, both for viscous and viscoelastic fluids. The correlation functions typically decay over several orders of magnitude in the considered time range,\cite{huan:12} and the MPC-intrinsic hydrodynamic fluctuations need to be averaged out. Nevertheless, good agreement is obtained between theory and simulations.

The qualitative different behavior in Fig.~\ref{fig:corr_dumbbell_sim_theory}(a) and (b) is in agreement with the theoretical expectations discussed in Sec.~\ref{sec:dumbbell}, since the radicand in Eq.~\eqref{eq:lambda_db} is positive for $l=1$  and negative for $l=\sqrt{3}$. Hence, for $l=1$, the correlation function decays exponentially according to Eq.~\eqref{eq:VACFFTT-dimer-damped}, whereas oscillations occur for longer bonds corresponding to Eq.~\eqref{eq:VACFFTT-dimer-simple-root}.
By fitting the theoretical expressions \eqref{eq:VACFFTT-dimer-simple-root} and \eqref{eq:VACFFTT-dimer-damped}, respectively, we find the relaxation time $\tau_1 \approx 2.8 \, l^2$ (see also Ref.~\onlinecite{kowa:13}).
This value agrees reasonably well  with the theoretical prediction $3.1 \, l^2$ following from the relaxation time Eq.~\eqref{eq:relax_time_disc} with the friction coefficient $\gamma=6\pi \eta R_H$,
where the hydrodynamic radius of a monomer  is $R_H=0.113$.\cite{kowa:13}

Evidently, our simulations and the theoretical approach yield long-range hydrodynamic correlations. The emergence of such correlations is not unexpected, since both the MPC simulations and
the (generalized) Navier-Stokes equations conserve momentum. For the relatively short polymer chains, Rouse-like
relaxation can be expected, because Zimm-type hydrodynamics requires long polymers, while the dumbbell relaxation time is only weakly affected by ``fluid'' correlations.\cite{kowa:13}

\subsubsection{Decamer Fluid ($N=10$)}

\begin{figure}[t!]
\begin{center}
\includegraphics[width=\columnwidth]{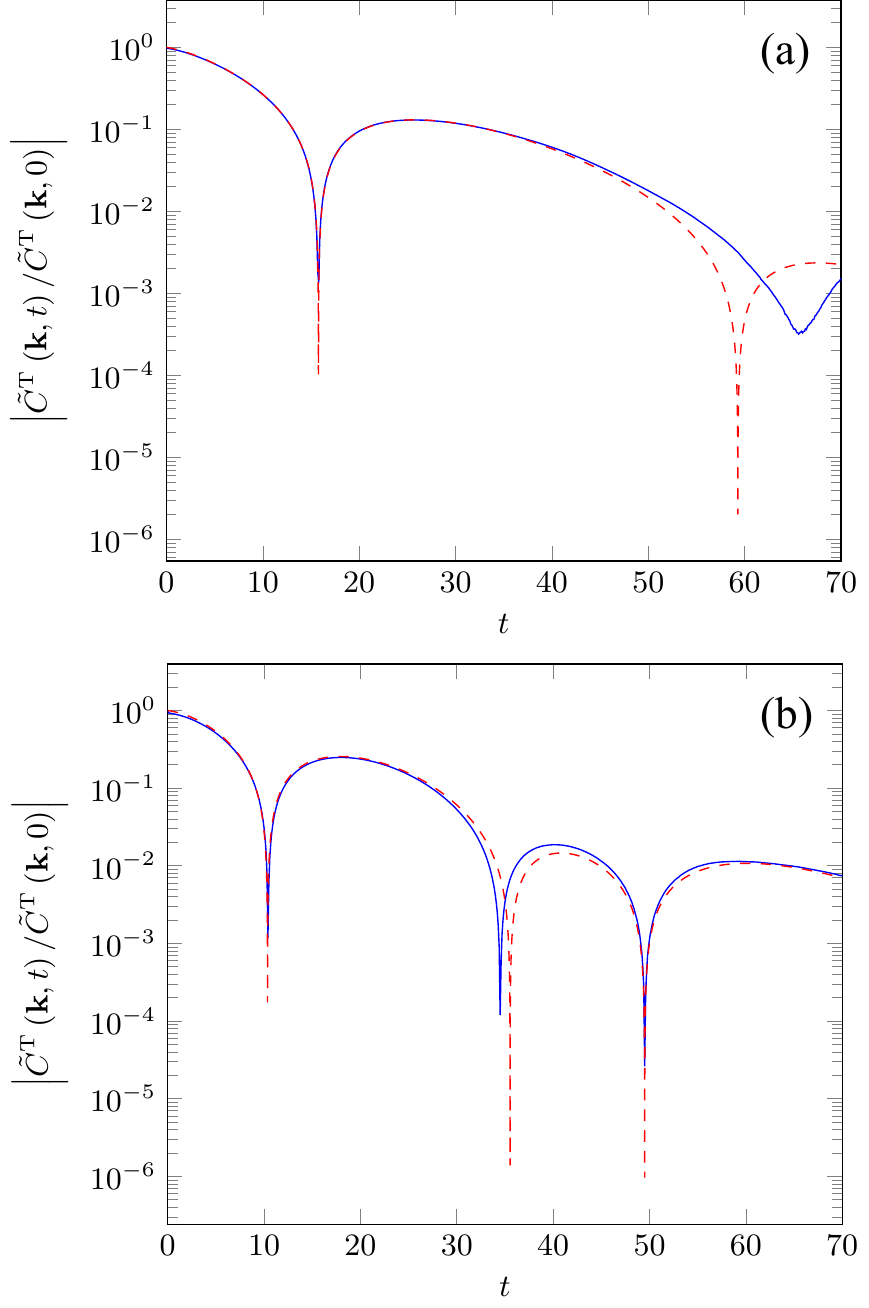}
\end{center}
\caption{Transverse velocity autocorrelation functions of decameres  ($N=10$) as a function of time $t$ from simulations (solid) and analytical theory (dashed) for the bond length (a) $l=1$ ($K=3$) and (b) $l=\sqrt{3}$ ($K=1$).
The wave vector is $\bm k = (2\pi/30,0,0)^T$. As in Fig.~\ref{fig:corr_dumbbell}, the correlation functions assume negative values (even loops). The theoretical curves are fitted to the simulation data.
} \label{fig:corr_decamer_sim_theory}
\end{figure}

Figure~\ref{fig:corr_decamer_sim_theory} presents simulation and theoretical results
of $\tilde C^T(\bm k,t)$ for decamers of various bond lengths.
Fitting of Eq.~\eqref{VACFFTT-general-solution-simple-roots} to the simulations data yields the relation  $\tau_1 \approx 54 \, l^2$ for the bond-length dependence of the longest relaxation time.
The theoretically predicted value $\tau_1 \approx 63 \, l^2$,  according to Eq.~\eqref{eq:relax_time_disc}, is somewhat larger, when the hydrodynamic radius $R_H=0.113$ is used.\cite{kowa:13}
The relaxation times are,  compared to a dumbbell fluid, longer and only damped oscillating correlation functions occur for the considered  $\bm k$ vectors, as expected theoretically.
The comparison of Fig.~\ref{fig:corr_decamer_sim_theory}(a) and (b) indicates an increase in the frequency with increasing relaxation time, in agreement with the theoretical expectations. Moreover, the evident different time intervals between zeros of $\tilde C^T(\bm k,t)$ in Fig.~\ref{fig:corr_decamer_sim_theory}(b) reflect the presence of multiple relaxation times.

As for the dumbbell fluid, we find very good agreement between the simulation data and the theoretical prediction over the presented time window.
In general, our results emphasize the strongly correlated polymer dynamics by the momentum-conserving interaction, i.e., long-range hydrodynamics.

\subsection{Center-of-Mass Velocity Correlation Function in Real Space}

The autocorrelation function of the center-of-mass velocity of a polymer in real space, Eq.~\eqref{eq:corr_cm_def}, can directly be calculated. For a compressible fluid, like the MPC fluid, $C_{cm}(t)$ comprises contributions from transverse and
longitudinal modes, which  cannot simply by extracted from the correlation function \eqref{eq:corr_cm_def} determined in simulations. However, the longitudinal modes affect the short-time behavior of $C_{cm}(t)$ only, since the
longitudinal correlation function decays exponentially,\cite{huan:12,huan:13} and
the longer-time hydrodynamic properties are determined by the transverse correlation function with its long-time tail. Hence, the correlation function $C_{cm}(t)$ of the MPC fluid exhibits the correct
long-time behavior.
Moreover, the short-time behavior of the correlation function reflects the partitioning of space into
collision cells of the MPC approach. Hydrodynamics appears only on length scales larger than the lattice
constant $a$ of the collision-cell lattice.\cite{huan:12} Consequently, at short times, $t \lesssim 5$ (Fig.~4),
the simulation results deviate from the solution of the continuum Navier-Stokes equations, independent of
polymer length, as is illustrated in Refs.~\onlinecite{huan:12,huan:13,pobl:14} for various systems.
Therefore, agreement between theory and simulations can only be expected at longer times.
This does not affect the dynamics of embedded colloids or polymers, because it is determined by the
hydrodynamic long-time tail.

\begin{figure}[t!]
\begin{center}
\includegraphics[width=\columnwidth]{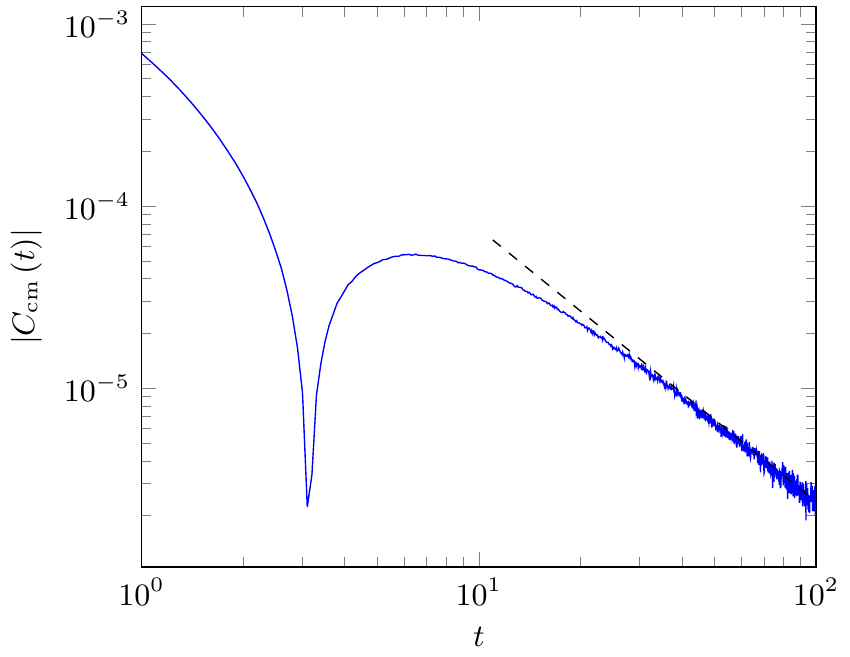}
\end{center}
\caption{Time dependence of the velocity autocorrelation function for  polymers with $N=100$ monomers and $l=1$ ($K=3$). The size of the simulation box is  $L_S =100$. The dashed line indicates the power law $t^{-3/2}$.
} \label{fig:corr_time}
\end{figure}

The correlation function for polymers of length $N=100$ and bond length $l=1$ is presented in Fig.~\ref{fig:corr_time}.
At $t=0$,  $C_{\textrm{cm}}(0) = 3k_BT/mN$ according to the equipartition theorem. (Note that this value includes transverse and longitudinal contributions.)
For short times, i.e., for $t \lesssim  1$,
$C_{\textrm{cm}}(t)$ reflects the discrete-time MPC procedure, with the first MPC collision at $t=h=0.1$. For $t\gtrsim 3$, the correlation function becomes negative due to viscoelasticity, as discussed in Sec.~\ref{sec:corr_cm_cont},  Eq.~\eqref{eq:corr_cm_cont}.
At longer times, the correlation function decays in
a power-law manner as $|C_{\textrm{cm}}(t)| \sim t^{-3/2}$, in agreement with the theoretical expectation for the regime  $\tau_l \approx 1\ll t  \ll \tau_R \approx 6 \times 10^3$ (cf. Sec.~\ref{sec:com_corr}). In the simulations, we did not reach the asymptotic value for $t \to \infty$ (Eq.~\eqref{eq:corr_asymp_t}), where the correlation function is positive again. The dependence $t^{-3/2}$ of the correlation function emphasizes and reflects the relevance of hydrodynamic interactions in a viscoelastic fluid of phantom polymers.

\section{Conclusions} \label{sec:conclusions}

We have presented a polymer-based model for a viscoelastic fluid and its implementation in a multiparticle collision dynamics algorithm.
The fluid properties have been characterized by the transverse velocity autocorrelation function. The comparison between analytical predictions, based on the Navier-Stoks equation, and simulation results shows good agreement and, thus, confirms the suitability of applied implementation to describe viscoelastic fluids.

Polymer elasticity strongly affects the velocity correlation function $\tilde C^T(\bm k,t)$, and leads to damped oscillations over a certain range of wave vectors. However, for long times and large length scales ($|\bm k| \to 0$), we expect and predict an exponential decay of $\tilde C^T(\bm k,t) \sim e^{- \nu_f \bm k^2 t}$  as function of time, with the kinematic viscosity of the polymeric fluid.
This implies a  long-time tail for the polymer center-of-mass velocity correlation function $C_{\textrm{cm}}(t) \sim t^{-3/2}$. On these scales, the polymer melt behaves as a fluid in terms of the correlation functions and, hence,  exhibits hydrodynamics with the effective viscosity $\eta_f$.
More interestingly, for very long polymers an additional power-law time regime can be identified.  In the range $\tau_l < t < \tau_R$ between the relaxation time on the scale of a bond and the Rouse time of the whole polymer, with a nearly continuous mode spectrum,  $\tilde C^T(\bm k,t)$ is negative and exhibits the power-law dependence $t^{-3/2}$ rather than an exponential decay with time. Fourier transformation of the correlation function weighted by static structure factor maintains the time dependence, such that the polymer center-of-mass correlation function in real space shows the same time dependence. This rather distinct  behavior is a consequence of the wide spectrum of modes, and thus, is a polymer-specific property. It reflects a strong influence of the polymer internal dynamics on the overall hydrodynamic behavior of the fluid.\cite{huan:13}

Here, we have only considered phantom polymers. Extension to  polymers with excluded-volume interactions would be interesting, specifically regarding the impact of excluded-volume interactions  on the correlation functions.
Yet, the results of Ref.~\onlinecite{fara:12.1} show that  even then the correlation function of a non-entangled polymer melt exhibits a long-time tail with the decay $t^{-3/2}$ in three dimensions.

Our results on the presence of a hydrodynamic long-time tail are consistent with theoretical predictions for phantom polymers. \cite{free:81} However, as pointed out in Ref.~\onlinecite{fara:12.1}, the studies of self-avoiding polymers seem to contradict the widely accepted view that hydrodynamic interactions are screened in polymer melts. \cite{doi:86} To be precise, the statement is typically used in the context of polymer solutions, where polymers are dissolved in a fluid, and momentum conservation is assumed to be violated for the fluid due to the immobile polymer matrix. Studies of the dynamical interplay of polymers and fluid would be desirable for a better understanding of screening. The presented simulation approach of polymers and fluid are very well suited for such an endeavor.

\begin{acknowledgments} D.~T. thanks the Institute of Complex Systems for its hospitality during an extended visit.
G. K. and D. T. acknowledge financial support by the
Austrian Science Fund FWF within the SFB ViCoM
(F41).   D.T. acknowledges financial support by the FWF under Proj.
No. I3846-N36, as well as computing time granted  by the Vienna Scientific
Cluster.  R. G. W. and G. G. gratefully acknowledge the computing time granted through JARA-HPC on the supercomputer JURECA at Forschungszentrum J\"ulich.
\end{acknowledgments}

\appendix

\section{General Inverse Laplace Transform of $\hat C^T(\bm k,s)$}
\label{app:Inverse-Laplace-Transform-of-VACFFTTLT}

To calculate the inverse Laplace transform $\tilde C^T(\bm k,t)$ of $\hat C^T(\bm k,s)$ of Eq.~\eqref{Rouse-VACFFTTLT}, the denominator of the right-hand side,
\begin{align}
D(s) = \varrho s + \bm k^2 \left(\eta + \varphi k_BT \sum_{p=1}^{N-1} \left(s + 2/ \tau_n \right)^{-1} \right) ,
\end{align}
is multiplied by $W(s) = \prod_{p=1}^{N-1} (s + 2/\tau_n)$, which yields
\begin{align}
P(s) = & \ D(s) W(s) \\ \nonumber  = & \
(\varrho s + \eta \bm k^2)W(s)+ \bm k^2 \varphi k_BT \sum_{n=1}^{N-1}  \prod_{\substack{p=1 \\ p\neq n}}^{N-1} (s + 2/\tau_n) ,
\end{align}
a polynomial in $s$ of degree $(N-1)$.
Let $P_n$, $n=1,\ldots,M$, be the $M$ distinct roots of $\fluidDensity^{-1} P \left( s \right)$,
each of multiplicity $M_n$,
so that $\sum_{n=1}^M M_n = N$, hence,
$P(s)= \varrho \prod_{m=1}^{M}\left(s-P_m \right)^{M_m}$.
For $W\!\left( s \right)$,
which is also a polynomial in $s$,
let $W_n$ be the coefficient of $s^n$,
so that $W(s) = \sum_{m=0}^{N-1} W_n s^n $.
Then, \eqnref{Rouse-VACFFTTLT} can be written as
\begin{align} \nonumber
\hat C^T(\bm k,s)  & \ = \varrho \tilde C^T(\bm k,0) \frac{W(s)}{P(s)} \\
& \ = \tilde C^T(\bm k,0) \sum_{n=0}^{N-1} \frac{W_n s^n}{\prod_{m=1}^M \left( s - P_m \right)^{M_m}} .
\end{align}
Inverse Laplace transformation of the terms ${s^n \prod_{m=1}^M (s - P_m)^{-M_m}}$ yields \cite{Erdelyi1954,Prudnikov1992} 
\begin{align} \label{VACFFTT-general-solution}
\frac{\tilde C^T(\bm k,t)}{\tilde C^T(\bm k,0)} =  \sum_{n=0}^{N-1} W_n \sum_{m=1}^M e^{P_mt} \sum_{l=1}^{M_m}
\frac{A_{nml}(P_m) t^{M_m-l}}{(M_m-l)!(l-1)!} ,
\end{align}
with
\begin{align}
A_{nml}(x)=\frac{d^{l-1}}{dx^{l-1}} \left(x^n \prod_{\substack{j=1 \\ j\neq m}}^M \left(x-P_j\right)^{-M_j} \right) .
\end{align}
Note that the degree of the polynomial of the enumerator  is higher than the one of the denominator.
If $P\!\left(s\right)$ only has simple roots,
i.e., $M_n=1$ for all $n$, Eq.~\eqref{VACFFTT-general-solution} simplifies to
\cite{ober:12,
Erdelyi1954} 
\begin{align} \label{VACFFTT-general-solution-simple-roots}
\frac{\tilde C^T(\bm k,t)}{\tilde C^T(\bm k,0)} =  \sum_{n=0}^{N-1} W_n \sum_{m=1}^N P_m^n e^{P_mt}
\prod_{\substack{j=1 \\ j\neq m}}^N \left(P_m-P_j\right)^{-1} .
\end{align}

\section{Inverse Laplace Transformation for Continuous Polymer} \label{app:inv_laplace}

The correlation function \eqref{eq:corr_cont_poly_s} is of the form
\begin{align}
\hat f (s) = \frac{1}{s+b/\sqrt{s}} = \frac{\sqrt{s}}{s^{3/2}+b} .
\end{align}
The inverse Laplace-transform  $f(t) =  {\cal{L}}^{-1}[\hat f(s);t]$ can be obtained  from that of $\hat H(\sqrt{s})$,
defined as
\begin{align}
\hat H(s) = \frac{s}{s^3 +b} ,
\end{align}
according to \cite{ober:12}
\begin{align} \nonumber
f(t)  = & \  {\cal{L}}^{-1}[\hat f(s);t] =  {\cal{L}}^{-1}[\hat H(\sqrt{s});t] \\ \label{app:inv_laplace_integral} =  & \ \frac{1}{2 \sqrt{\pi t^3}} \int_0^{\infty} \tau \exp{-\frac{\tau^2}{4 t}} H(\tau) d\tau .
\end{align}
Partial fraction decomposition of $s^3 +b$ in $\hat H(\sqrt{s})$
and straightforward inverse Laplace  transformation yields
\begin{align} \label{eq:H}
H(t) = \frac{1}{3\sqrt[3]{b}}\left( e^{\kappa t} \cos(\omega t) + \sqrt{3} e^{\kappa t} \sin(\omega t) - e^{-2 \kappa t} \right) ,
\end{align}
with the abbreviations $\kappa = \sqrt[3]{b}/2$ and $\omega = \sqrt{3} \sqrt[3]{b}/2$.
Evaluation of the integral \eqref{app:inv_laplace_integral} gives,  with \eqref{eq:H},
\begin{align} \nonumber
f(x) =  \frac{1}{3} & \left\{\mathrm{erfcx}\left(-\frac{1}{2}[1 + i \sqrt{3}]\sqrt{x} \right)  \right. \\ \label{app:int_laplace}  & \left. +  \ \mathrm{erfcx}\left(-\frac{1}{2}[1 - i \sqrt{3}]\sqrt{x}\right)
+ \  \mathrm{erfcx}\left( \sqrt{x}\right)  \right\} ,
\end{align}
with $x =b^{2/3} t$.
Here, $\mathrm{erfcx}(y)$ is the scaled complementary error function
\begin{align}
\mathrm{erfcx}(y) =  e^{y^2} \left(1 - \frac{2}{\sqrt{\pi}}\int_0^y e^{- u^2} du \right) .
\end{align}
We like to mention that Laplace transformation of Eq.~\eqref{eq:corr_cont_poly_s} including $\eta$ can be performed in a similar manner.

\section{Center-of-Mass Velocity Autocorrelation Function---Connection to Dynamic Structure Factor} \label{app:com_corr}
The polymer center-of-mass correlation function is given by (cf. Eq.~\eqref{eq:corr_cm_def})
\begin{align}  \label{app:corr_cm_def}
C_{\textrm{cm}}(t)= \frac{1}{N^2} \sum_{i=1}^N \sum_{j=1}^N
\lla \bm v(\bm r_i,t) \cdot \bm v(\bm r_j,0) \rra .
\end{align}
The Fourier representation
\begin{align}
\bm v (\bm r_i,t) = \frac{1}{(2 \pi)^2} \int \tilde{\bm v}(\bm k,t) e^{i \bm k \cdot \bm r_i(t)} d^3k
\end{align}
yields
\begin{align}
&C_{\textrm{cm}}(t)=   \frac{1}{N^2 (2\pi)^2} \\ \nonumber &
 \times \sum_{i=1}^N \sum_{j=1}^N
\lla \tilde {\bm v}(\bm k,t) \cdot \tilde {\bm v}(\bm k',0) \rra  e^{i \bm k \cdot \bm r_i(t)} e^{i \bm k' \cdot \bm r_j (0)} d^3k d^3k' .
\end{align}
With the definition of the fluid correlation function $\langle \tilde {\bm v}(\bm k,t) \cdot \tilde {\bm v}(\bm k',0) \rangle \sim \delta(\bm k + \bm k') \tilde C(\bm k,t)$ \eqref{eq:def_velcor}, and the dynamic structure factor \eqref{eq:dyn_struct_gen}
\begin{align}
 S(\bm k,t) = \frac{1}{N} \sum_{i=1}^N \sum_{j=1}^N \langle e^{i \bm k \cdot (\bm r_i(t) - \bm r_j(0))} \rangle ,
\end{align}
Eq.~\eqref{app:corr_cm_def} becomes
\begin{align}
C_{\textrm{cm}}^T(t) = \frac{1}{(2\pi)^3N} \int S(\bm k,t) {\tilde C}^T(\bm k,t) d^3 k ,
\end{align}
which is Eq.~\eqref{eq:com_corr}.

\end{document}